\documentclass[amsmath,amssymb,aps,floatfix]{revtex4-2}

\usepackage{graphicx}
\usepackage{dcolumn}
\usepackage{bm}
\usepackage[mathscr]{euscript}
\usepackage{mathrsfs}
\usepackage{enumerate}

\usepackage{tikzsymbols}
\usepackage{natbib}
\usepackage{float}

\usepackage{mathtools} 

\usepackage{tikz,xcolor,hyperref}

\definecolor{lime}{HTML}{A6CE39}
\DeclareRobustCommand{\orcidicon}{
	\begin{tikzpicture}
	\draw[lime, fill=lime] (0,0) 
	circle [radius=0.2] 
	node[white] {{\fontfamily{qag}\selectfont \tiny ID}};
	\draw[white, fill=white] (-0.0625,0.095) 
	circle [radius=0.007];
	\end{tikzpicture}
	\hspace{-2mm}
}

\foreach \x in {A, ..., Z}{\expandafter\xdef\csname orcid\x\endcsname{\noexpand\href{https://orcid.org/\csname orcidauthor\x\endcsname}
			{\noexpand\orcidicon}}
}

\begin{document}


\title{Gauge and Poincare properties of the UV cutoff and UV completion in quantum field theory}

\author{Durmu{\c s} Demir\orcidA{}}
 \email{durmus.demir@sabanciuniv.edu}
 \homepage{http://myweb.sabanciuniv.edu/durmusdemir/}
\affiliation{Sabanc{\i} University, Faculty of Engineering and Natural Sciences, 34956  {\.I}stanbul, Turkey}

\date{\today}

\begin{abstract}
The ultraviolet (UV) cutoff on a quantum field theory (QFT) can explicitly break or conserve the Poincare (translation) symmetry. And the very same cutoff can explicitly break or conserve the gauge symmetry. In the present work, we perform a systematic study of the UV cutoff in regard to its gauge and Poincare properties, and construct UV completions restoring the broken gauge symmetry. In the case of Poincare-conserving UV cutoff, we find that the gauge symmetry gets restored via the Higgs mechanism. In the case of Poincare-breaking UV cutoff, however, we find that the flat spacetime affine curvature takes the place of the Higgs field and, when taken to curved spacetime, gauge symmetry gets restored at the extremum of the metric-affine action. We also find that gravity emerges at the extremum if the QFT under concern consists of new particles beyond the known ones. The resulting emergent gravity plus renormalized QFT setup has the potential to reveal itself in various astrophysical, cosmological and collider phenomena.
\end{abstract}

\maketitle

\tableofcontents

\section{Introduction}
In general, QFTs can possess  momentum cutoffs in the IR (like the  confinement scale $\Lambda_{QCD}$) or in the UV (like the gravitational scale $M_{Pl}$). Fundamentally, QFTs rest on a classical action and a certain form of UV cutoff. Low-energy effective QFTs develop power-law and logarithmic sensitivities to the UV cutoff \cite{ghp1,weinberg-eft}. In general, stronger the sensitivity to the cutoff stronger the destabilization of the QFT from its natural scale at the IR. In this sense, it is the strength of the sensitivity to the cutoff that decides  if the QFT remains natural or not  under  quantum fluctuations. In the pre-Higgs era, alleviation of the UV sensitivity was the reason for and requirement from all models of new physics beyond the standard model (SM). One such model, softly-broken weak-scale supersymmetry \cite{dimopoulos}, was introduced to alleviate UV sensitivity  via the spin-statistics theorem \cite{susy}. In the post-Higgs era, alleviation of the  UV sensitivity has become a much more perplexing problem \cite{giudice-0,williams,wells-0,dine} as because experiments at the LHC have detected no new particles at the weak scale \cite{ATLAS,CMS}. In spite of various interesting mechanisms proposed so far \cite{craig}, alleviation of the UV sensitivity is a thorny open problem and calls for a resolution.  

The goal in the present work is to complete the effective QFT in the UV insofar as the
symmetry structure of the UV cutoff permits. If the UV cutoff is the mass of a heavy field then it preserves the Poincare symmetry and we call it Poincare-conserving UV cutoff. It will in general generate quadratic divergent corrections to scalar masses. If the heavy field is a vector field then divergences can be absorbed by adding massive scalars restoring the gauge symmetry in the massive vector field as in the usual Higgs mechanism. In this case, as will be analyzed in Sec. II, the UV divergent terms get transmuted into logarithms and powers of the masses of the added heavy scalars.

On the other hand, if the UV cutoff is not the mass of a field (a hard momentum cutoff like the gravitational scale) then it breaks the Poincare symmetry. It will generate via loops divergent corrections to scalar, fermion and gauge boson masses.  This effect of the cutoff breaks gauge symmetry. Conventionally, this effect is dealt with using the dimensional regularization, which preserves the gauge symmetry since loop-induced gauge boson mass terms to cancel automatically. Here, in this work, we seek a more physical UV completion of the effective QFT. In this regard, our proposal is to absorb the vector boson mass divergences by coupling the vector bosons to a possible affine curvature. This structure naturally restores the gauge symmetry broken by the UV cutoff, but at the price of coupling the gauge theory to space-time curvature. Following this route to its logical conclusion, in Sec. III we find that the UV completion generates a pure gravity action, that is, emergent gravity.

This UV completion mechanism does not solve the various hierarchy or fine-tuning problems
of the SM plus gravity, but it allows us to tackle these problems in a new way.
It also has implications that we find interesting for the physics beyond the SM.
It requires the presence of beyond-the-SM (BSM) particles to generate a proper gravitational action.  
The SM Higgs sector is not necessarily destabilized by the heavy BSM particles since SM-BSM interaction is not a necessity. As discussed in Sec. III, these BSM particles can have observable effects in, for instance, black holes, early Universe and high-luminosity colliders.

The UV completion is constructed in accordance with the gauge and Poincare properties of the UV cutoff. The details of the construction, section by section, can be summarized as follows: As already mentioned, Poincare-conserving UV cutoff (denoted as $M_\wp$ from now on)  is the mass of a field namely it is the Casimir invariant of the Poincare group \cite{bogolyubov}. The Poincare-breaking UV cutoff (denoted as $\Lambda_\wp$ from now on), on the other hand, is like the confinement scale or the gravitational scale namely it is not the mass of a particle. These two kinds of UV cutoffs imply different UV sensitivities, different alleviation mechanisms and different UV completions. And all these properties will be studied in detail in  Sec.  \ref{Poincare-Conserving UV Cutoff} for the Poincare-conserving UV cutoff $M_\wp$ and in Sec. \ref{Poincare-Breaking UV Cutoff} for the Poincare-breaking UV cutoff $\Lambda_\wp$. The resulting UV completions  will be contrasted in Sec.  \ref{Poincare-Conserving vs Poincare-Breaking UV Cutoffs}. 

In principle, both the Poincare-conserving and Poincare-breaking UV cutoffs can break gauge symmetries. Indeed, the Poincare-conserving UV cutoff $M_\wp$ conserves gauge symmetries if it is the mass of a scalar or a suitable fermion (as  will be discussed in Sec. \ref{Gauge-Conserving UV Cutoff} in the dimensional regularization \cite{dim-reg1,dim-reg2,regularization}). But the same  Poincare-conserving UV cutoff $M_\wp$ breaks gauge symmetries if it is the mass of a vector field (as will be discussed in Sec. \ref{Gauge-Breaking UV Cutoff} with the examples of weak bosons \cite{IVB1,IVB2} and Pauli-Villars fields \cite{PV-0,PV-1,PV-2}). The Poincare-breaking UV cutoff $\Lambda_\wp$ \cite{gauge-break1, cutoff1}, on the other hand, breaks all the gauge symmetries \cite{gauge-break2,gauge-break3,gauge-break4}. In general, gauge symmetry breaking is caused by the cutoff-induced gauge boson masses and, in the philosophy of the Higgs mechanism, gauge symmetries could be restored by promoting the UV cutoff to suitable dynamical fields  \cite{review1,review2,review3,review4}. For the Poincare-conserving UV cutoff $M_\wp$ in Sec. \ref{Gauge-Breaking UV Cutoff}, the said dynamical field turns out to be a Higgs scalar so that the massive vector gets completed into a gauge field via the Higgs mechanism. For the Poincare-breaking UV cutoff $\Lambda_\wp$ in Sec. \ref{Gauge Symmetry Restoration at the Extremum of Action}, on the other hand,  the said dynamical field turns out to be the affine curvature (not the metrical curvature) \cite{affine1,affine2,affine3,biz-beyhan}. In Sec. \ref{Detached Regularization},  detached regularization is introduced as a new regularization framework in which  while the logarithmic divergences involve the usual renormalization scale $\mu$ the power-law divergences go with the UV cutoff $\Lambda_\wp$ \cite{bizimki}. In Sec. \ref{Effective QFT}, flat spacetime effective QFT is formed in the detached regularization. In Sec. \ref{Gauge Symmetry Restoration at the Extremum of Action}, it is shown that the UV cutoff can be consistently promoted to flat spacetime affine curvature in the same philosophy as the promotion of the UV cutoff to Higgs scalar in Sec. \ref{Gauge-Breaking UV Cutoff}. In Sec. \ref{Effective QFT in Curved Spacetime}, flat spacetime effective QFT is carried into spacetime of a curved metric such that it is found that, in an effective QFT, curvature terms can arise only in the gauge sector. In Sec. \ref{Affine Dynamics: From UV Cutoff to IR Curvature}, affine curvature is integrated  out and it is found that at the extremum of the affine action all gauge symmetries get restored and general relativity (with quadratic curvature terms) emerge. The physics implications of the emergent gravity is discussed in Sec. \ref{Salient Properties of the Emergent Gravity}.  

In Sec. \ref{Poincare-Conserving vs Poincare-Breaking UV Cutoffs},  a comparative discussion is given of the UV completion mechanisms for the Poincare-conserving and Poincare-breaking UV cutoffs.

The work is concluded in Sec. \ref{Conclusion}. 

\section{Poincare-Conserving UV Cutoff}
\label{Poincare-Conserving UV Cutoff}
The UV cutoff $M_\wp$ conserves the Poincare symmetry if it is the mass of a particle. It conserves because the cutoff itself is  a Casimir invariant of the Poincare group and it can give cause therefore to no Poincare breaking \cite{bogolyubov}. Field theoretically, all one has is a heavy field of mass $M=M_\wp$ setting the UV boundary of the QFT. This heavy field can be a physical field or an unphysical (regulator) field. In the former, the heavy field is just one of the fields in the QFT and dimensional regularization is the natural regularization scheme. In the latter, the heavy field is originally not a part of the QFT as it is added as a regulator for UV divergences, and Pauli-Villars regulators \cite{PV-0} set a good example of such unphysical fields. These fields soften the UV divergences via their wrong-sign propagators coming from their wrong-sign kinetic and mass terms  (negative-norm Lagrangian)  \cite{PV-1,PV-2}. In  Sec. \ref{Gauge-Conserving UV Cutoff} and  Sec. \ref{Gauge-Breaking UV Cutoff} below our analysis will focus on solely the heavy physical fields in view of the naturalness problem in the SM. However, as will be seen in Sec. \ref{Gauge-Breaking UV Cutoff}, our methodology can be directly extended to Pauli-Villars fields. 

\subsection{Gauge-Conserving UV Cutoff}
\label{Gauge-Conserving UV Cutoff}
One example of Poincare- and gauge-conserving UV cutoff is the  mass $m_{\tilde \phi}=M_\wp$ of a heavy physical scalar ${\tilde \phi}$. (One other example would be the Pauli-Villars scalar regulators \cite{PV-0, PV-1,PV-2}.)  Keeping all the terms not forbidden by symmetries, inclusion of ${\tilde \phi}$ 
extends the QFT action $S[\eta,\digamma]$ as  
\begin{eqnarray}
S[\eta,\digamma] \longrightarrow S[\eta,\digamma] +\int d^4 x \sqrt{-\eta} \left\{(D_\mu {\tilde \phi})^\dagger (D^\mu {\tilde \phi}) - M_\wp^2 {\tilde \phi}^\dagger {\tilde \phi} -\lambda_{\tilde \phi} ({\tilde \phi}^\dagger {\tilde \phi})^2 
+{\mathcal{L}}_{int}(\eta,\digamma;{\tilde \phi})\right\}
\label{DR-1}
\end{eqnarray}
in which $\eta_{\mu\nu}$ is the flat Minkowski metric,  $\digamma\equiv \left\{\phi,\psi,V_\mu\right\}$  is a collective  label for scalars $\phi$, fermions $\psi$ and gauge bosons $V_\mu$ in the QFT, $D_\mu {\tilde \phi}$ is the gauge-covariant derivative of ${\tilde \phi}$, and ${\mathcal{L}}_{int}(\eta,\digamma;{\tilde \phi})$ encodes interactions between  the QFT fields $\digamma$ and the heavy scalar ${\tilde \phi}$. By construction, this extended action conserves both the Poincare and gauge symmetries. 

The extended QFT in (\ref{DR-1}) has no UV boundary, that is, there is no  finite cutoff scale on the loop momenta $\ell_\mu$. A typical loop amplitude
\begin{eqnarray}
I_{n}(\mu) &=& \int \frac{d^4 \ell}{(2\pi)^4} \frac{1}{(\ell^2 - m^2 +i0)^n}
\label{DR-20}
\end{eqnarray}
can therefore be computed using the  dimensional regularization scheme \cite{dim-reg1,dim-reg2,regularization}
\begin{eqnarray}
I_{n,D}(\mu) &=& \mu^{4-D}\int \frac{d^D \ell}{(2\pi)^D} \frac{1}{(\ell^2 - m^2 +i0)^n}
= \left\{\begin{array}{ll} 
\frac{i m^2}{16 \pi^2} \left(1 + \log \frac{\mu^2}{m^2}\right) & \quad n=1, D=4\\ 
\frac{i}{16 \pi^2} \log \frac{\mu^2}{m^2} & \quad n=2, D=4
\end{array}\right.
\label{DR-2}
\end{eqnarray}
after subtracting away $1/(D-4)$ poles at the 
scale $\mu$ in the modified minimal subtraction (${\overline{MS}}$) scheme  \cite{MSbar-1,MSbar-2}.  In this framework, integration of  ${\tilde \phi}$ out of the spectrum  modifies the QFT action $S[\eta,\digamma]$ by the correction \cite{cutoff1,gauge-break2}
\begin{eqnarray}
\delta S[\eta, \digamma; M_\wp^2, \log\mu] =  \int d^4 x \sqrt{-\eta} \left\{c_{\phi} \lambda_{\phi} M_\wp^2 \log \frac{M_\wp^2}{\mu^2} \phi^\dagger \phi + c_{\psi} \lambda^2_{\psi} m_\psi \log \frac{M_\wp^2}{\mu^2} {\overline{\psi}} \psi + 0 \times {\rm tr} \left[V_\mu V^\mu\right] \right\}
\label{eff-ac-DR}
\end{eqnarray}
corresponding to renormalizable interactions $\lambda_\phi (\phi^\dagger \phi) ({\tilde \phi}^\dagger {\tilde \phi})$ and $[\lambda_\psi{\tilde \phi} \overline{\psi} \psi + {\rm h. c.}]$ in the 
interaction lagrangian ${\mathcal{L}}_{int}(\eta,\digamma;{\tilde \phi})$. This correction action reveals that  $\delta m_\phi^2 \propto m_{\tilde \phi}^2 \log \!\frac{m_{\tilde \phi}^2}{\mu^2}$ 
for scalars $\phi$,  $\delta m_\psi \propto m_\psi \log \!\frac{m_{\tilde \phi}^2}{\mu^2}$ for fermions $\psi$, and $\delta M_{V} = 0$ for gauge fields $V_\mu$ (with ${\rm tr}[\dots]$ standing for color trace). With these mass corrections, the loop-induced QFT action (\ref{eff-ac-DR}) possesses two important properties: 
\begin{enumerate}
    \item The action (\ref{eff-ac-DR}) involves no corrections like $M_\wp^2 \eta_{\mu\nu}$ in the gauge sector. This means that massless gauge fields remain massless namely gauge symmetries are strictly conserved.   

    \item Loop corrections can destabilize the scalar field sector. Indeed, gauge boson masses remain unshifted, fermion masses shift only logarithmically with $M_\wp$, and yet scalar masses shift  quadratically with $M_\wp$. This means that larger the UV cutoff $M_\wp$ (or, equivalently, mass $M_\wp$ of the heavy scalar ${\tilde \phi}$) larger the corrections  $M_\wp^2 \log \!\frac{M_\wp^2}{\mu^2}$ to mass-squareds of light scalars. 
\end{enumerate}
In practise, naturalness of QFTs are measured by their sensitivity to the 
Poincare- and gauge-conserving UV cutoff $M_\wp$.   It is with this type of UV cutoff that unnaturalness of light scalars like the Higgs boson was perceived \cite{susskind,ghp1,wells}, and natural UV completions like  softly-broken supersymmetry were introduced \cite{dimopoulos}. And it is with this type of  UV cutoff that the null results from the LHC experiments \cite{ATLAS,CMS} have been interpreted as defying naturalness as a physical criterion \cite{giudice-0,dine} in regard to the hierarchy problem caused by the $\delta m_\phi^2 \propto M_\wp^2$ correction  \cite{lhp1,lhp2}.

\subsection{Gauge-Breaking UV Cutoff}
\label{Gauge-Breaking UV Cutoff}
One example of Poincare-conserving but gauge-breaking UV cutoff is the mass $M_{\tilde V}=M_\wp$ of a heavy physical vector ${\tilde V}_\mu$.
This field can be incorporated into  the QFT action $S[\eta,\digamma]$ via the change
\begin{eqnarray}
S[\eta,\digamma] \longrightarrow S[\eta,\digamma] +\int d^4 x \sqrt{-\eta} \left\{ -\frac{1}{4}   {\rm tr}\left[{\tilde V}_{\mu\nu} \eta^{\mu\alpha} \eta^{\nu\beta} {\tilde V}_{\alpha\beta}\right] + \frac{1}{2} M_\wp^2\, {\rm tr}\left[{\tilde V}_\mu  \eta^{\mu\nu} {\tilde V}_\nu \right]
+ {\mathcal{L}}_{int}(\eta,\digamma; {\tilde V}_\mu)
\right\}
\label{PV-1}
\end{eqnarray}
in which ${\tilde V}_{\mu\nu}$ is the field strength tensor of the vector  ${\tilde V}_\mu$ and  ${\mathcal{L}}_{int}(\eta, \digamma;{\tilde V}_\mu)$ is the interaction lagrangian between the QFT fields $\digamma$ and the vector ${\tilde V}_\mu$. One example of ${\tilde V}_\mu$ concerns physical massive vectors like, for example, the intermediate vector boson in Fermi theory (weak gauge bosons) \cite{IVB1,IVB2}. (One other example would be Pauli-Villars vector regulators \cite{PV-0,PV-1,PV-2} introduced to cancel the UV divergences caused by the QFT fields $\digamma$.) In general, independent of the circumstances they arise, massive vector fields cause problems especially with the renormalizability and gauge invariance. Restoration of gauge symmetry is therefore a necessity. The fundamental restoration method is the Stueckelberg method \cite{stueckelberg0,stueckelberg1} according to which the mass term $M_\wp^2 {\tilde V}_\mu  {\tilde V}^\mu$ is made gauge-invariant by forming the gauge-invariant vector ${\tilde V}_\mu - \partial_\mu {\tilde \chi}$ with the introduction of a scalar field ${\tilde \chi}$ such that $\partial_\mu {\tilde \chi}$ transforms in the same way as ${\tilde V}_\mu$. This method runs into problems with renormalizability for non-Abelian vectors. It gives nevertheless hints of the Higgs mechanism as a renormalizable linear sigma model. In fact, Higgs mechanism generates the mass $M_{\tilde V}=M_\wp$ of the vector field from the vacuum expectation value (VEV) of the Higgs field. And it will be utilized below for restoring the gauge symmetry.

\subsubsection{Promoting Poincare-Conserving UV Cutoff to Higgs Field}
\label{Gauge Symmetry Restoration at the Extremum of Energy}

The goal is to restore the the gauge symmetry associated with the massive vector ${\tilde V}_\mu$ in the extended QFT in  (\ref{PV-1}). The gauge boson mass $M_{\tilde V}=M_\wp$ is a Poincare-conserving mass scale, and the starting point is to promote it to a Poincare-conserving field. In this regard, it is sufficient to introduce a scalar field $\Phi$ transforming in the fundamental of the gauge group, and transfigure the vector field  mass term in (\ref{PV-1}) as a scalar-vector interaction
\begin{eqnarray}
M_\wp^2 {\rm tr}\left[ {\tilde V}_\mu \eta^{\mu\nu}  {\tilde V}_\nu\right]
\longmapsto \Phi^\dagger {\tilde V}_\mu \eta^{\mu\nu} {\tilde V}_\nu  \Phi
\label{MV-map}
\end{eqnarray}
under which the Poincare-conserving cutoff 
$M_\wp$ gets promoted to the scalar field $\Phi$. The gauge symmetry is still broken.  The scalar $\Phi$ here is actually no more than a spurion \cite{spurion1,spurion2}. In fact, extremization of the action (\ref{PV-1}) with respect to $\Phi$ gives $\Phi=0$, which comes to mean that gauge symmetry is restored by just setting gauge boson mass  to zero. This is not a UV completion. In fact, it is necessary to make $\Phi$ a dynamical field and, to this end, generalization of the scalar-vector interaction in (\ref{MV-map}) to $\Phi$-kinetic term 
\begin{eqnarray}
\Phi^\dagger {\tilde V}_\mu \eta^{\mu\nu} {\tilde V}_\nu  \Phi
\longmapsto ({\tilde D}_\mu \Phi)^\dagger \eta^{\mu\nu} ({\tilde D}_\nu \Phi)
\label{MV-map2}
\end{eqnarray}
generates the requisite dynamical structure. For a renormalizable $\Phi$-potential of the form 
\begin{eqnarray}
V(\Phi^\dagger \Phi) = \mu_\Phi^2 \Phi^\dagger \Phi + \frac{\lambda}{2} \left(\Phi^\dagger \Phi\right)^2
\label{poten}
\end{eqnarray}
gauge symmetry gets spontaneously broken by the Higgs VEV $\langle \Phi^\dagger \Phi \rangle = -\mu_\Phi^2/\lambda$ if $\mu_\Phi^2 <0$. This spontaneous breaking leads to the QFT action $S[\eta,\digamma]$ in (\ref{PV-1}) provided that $-\mu_\Phi^2/\lambda \simeq M^2_\wp$ which leads to $M_{\tilde V}^2 \simeq M_\wp^2$. It is via this spontaneous breaking that the intermediate massive vector bosons \cite{IVB1,IVB2} were realized as the gauge bosons of the electroweak gauge symmetry \cite{review1,review2,review3,review4}. 

In an alternative take, the gauge symmetry may not be broken at all. Indeed, if $\lambda>0$ and $\mu_\Phi^2 >0$  then the scalar $\Phi$ develops a vanishing VEV
\begin{eqnarray}
\langle \Phi^\dagger \Phi \rangle = 0
\label{poten-min}
\end{eqnarray}
at which ${\tilde V}_\mu$ remains strictly massless ($M_{\tilde V}^2 = 0$).  This state corresponds to restoration of the gauge symmetry associated with the massive vector ${\tilde V}_\mu$. In general, this symmetric phase could be relevant if the goal is to kill the gauge boson mass $M_{\tilde V}$, and Pauli-Villars vector \cite{PV-1,PV-2} could be one such field.  In this gauge-conserving minimum the extended QFT action in (\ref{PV-1}) undergoes a further extension to take the form
\begin{eqnarray}
S[\eta,\digamma;{\tilde V},\Phi] = S[\eta,\digamma] + \!\!\int\!\! d^4 x \sqrt{-\eta} \left\{\! -\frac{1}{4}   {\rm tr}\!\left[{\tilde V}_{\mu\nu} \eta^{\mu\alpha} \eta^{\nu\beta} {\tilde V}_{\alpha\beta}\right] + ({\tilde D}_\mu \Phi)^\dagger \eta^{\mu\nu} ({\tilde D}_\nu \Phi) - V(\Phi^\dagger \Phi) + 
 {\mathcal{L}}_{int}(\eta,\digamma; {\tilde V}_\mu,\Phi)
\!\right\}
\label{PV-11}
\end{eqnarray}
as a result of the maps (\ref{MV-map}) and (\ref{MV-map2}). The potential energy  $V(\Phi^\dagger \Phi)$ and the interaction lagrangian ${\mathcal{L}}_{int}(\eta,\digamma; {\tilde V}_\mu,\Phi)$ are all gauge-invariant.  Now,  integration of the heavy scalar $\Phi$ out of the spectrum induces the action correction \cite{cutoff1,gauge-break2}
\begin{eqnarray}
\delta S[\eta, \digamma; \mu_\Phi^2, \log\mu] =  \int d^4 x \sqrt{-\eta} \left\{c_{\phi} \lambda_{\phi} \mu_\Phi^2 \log \frac{\mu_\Phi^2}{\mu^2} \phi^\dagger \phi + c_{\psi} \lambda^2_{\psi} m_\psi \log \frac{\mu_\Phi^2}{\mu^2} {\overline{\psi}} \psi + 0 \times {\rm tr} \left[V_\mu V^\mu\right] + 0 \times {\rm tr} \left[{\tilde V}_\mu {\tilde V}^\mu\right] \right\}
\label{eff-ac-DR-2}
\end{eqnarray}
after computing loop amplitudes as in  (\ref{DR-2}) using the dimensional regularization  \cite{dim-reg1,dim-reg2,regularization} and
${\overline{MS}}$ subtraction \cite{MSbar-1,MSbar-2}.  The mass corrections  from $\Phi$-loops correspond to the renormalizable interactions $\lambda_\phi (\phi^\dagger \phi) ({ \Phi}^\dagger {\Phi})$ and $\lambda_\psi
{\Phi} \overline{\psi} \psi$ contained in ${\mathcal{L}}_{int}(\eta,\digamma; {\tilde V}_\mu,\Phi)$ in (\ref{PV-11}). It is clear that mass shifts are  $\delta m_\phi^2 \propto \mu_\Phi^2 \log \!\frac{\mu_\Phi^2}{\mu^2}$ 
for scalars $\phi$,  $\delta m_\psi \propto m_\psi \log \!\frac{\mu_\Phi^2}{\mu^2}$ for fermions $\psi$, and $\delta M_{\tilde V} = 0$
for gauge fields. Then, the loop-induced QFT action (\ref{eff-ac-DR-2}) possesses two important properties: 
\begin{enumerate}
    \item Massless gauge fields ($V_\mu$ in QFT and ${\tilde V}_\mu$ after the map (\ref{MV-map})) remain massless. This means that gauge symmetries in QFT are conserved. This also means that the gauge symmetry associated with the massive vector ${\tilde V}_\mu$ in (\ref{PV-1}) is restored in the gauge-conserving vacuum in (\ref{poten-min}).

    \item In the gauge-conserving vacuum in (\ref{poten-min}) all that is needed is $\mu_\Phi^2 > 0$ and $\lambda>0$. There is thus nothing that forbids the low $\mu_\Phi$ regime ($\mu_\Phi \ll M_\wp$). In this regime, $\Phi$-loops do not have to destabilize the light scalars $\phi$ in QFT since the shift in their masses 
   $\delta m_\phi^2 \propto \mu_\Phi^2 \log \!\frac{\mu_\Phi^2}{\mu^2} \ll m_\phi^2$ remains small. 
\end{enumerate}
In summary, the UV physics (the Higgs scalar $\Phi$ in the action (\ref{PV-11})) introduced to restore the gauge symmetry associated with the massive vector ${\tilde V}_\mu$ leads to a UV completion of the QFT in which the little hierarchy problem \cite{lhp1,lhp2} does not have to be a problem.

\section{Poincare-Breaking UV Cutoff}
\label{Poincare-Breaking UV Cutoff}
There are scales in nature that are not particle masses. The confinement scale in QCD (an IR scale) and fundamental scale of gravity (a UV scale) are such scales. This kind of UV cutoff breaks the Poincare symmetry since it is not the mass of an elementary particle namely not a Casimir invariant of the Poincare group \cite{bogolyubov}. One familiar example is the Lorentz-invariant but translation-breaking cut $\Lambda_\wp$ on the loop momenta $\ell_\mu$:   $-\Lambda_\wp^2 \leq \ell_\mu \ell^\mu \leq \Lambda_\wp^2$. Under this cut, the effective QFT remains Lorentz-invariant but breaks explicitly all the gauge symmetries since each gauge boson acquires a mass proportional to $\Lambda_\wp$ \cite{gauge-break1,gauge-break2,gauge-break3,gauge-break4}. Besides this, mass-squareds of scalars develop quadratic sensitivity to the UV cutoff $\Lambda_\wp$ \cite{cutoff1,ghp1}. And vacuum energy develops both quartic and quadratic sensitivities to the cutoff scale \cite{akhmedov0,akhmedov}.  

In this section, a detailed study will be performed of the question of how gauge symmetries can be restored in the presence of a Poincare-breaking UV cutoff. 

\subsection{Detached Regularization}
\label{Detached Regularization}
Detached regularization is a new regularization method proposed in \cite{bizimki}. It extends the dimensional regularization \cite{dim-reg1,dim-reg2,regularization} to QFTs with a UV cutoff such that the power-law and logarithmic divergences can be treated separately and independently. In is this scheme, the generic loop amplitude in (\ref{DR-20}) is  restructured in a way gathering contributions of $D=0$ (quartic), $D=2$ (quadratic), and $D=4$ (logarithmic) dimensions \cite{bizimki}
\begin{eqnarray}
I_{n,D}(\Lambda_\wp,\mu) &=& \left[\frac{\left(\delta_{[D]0} + \delta_{[D]2}\right)}{(8\pi)^{2-n}} \Lambda_\wp^{4-2n} \mu^{2n-D} + \delta_{[D]4}\;  \mu^{4-D}\right]  \int \frac{d^D p}{(2\pi)^D} \frac{1}{(p^2-m^2+i0)^n} \label{int-Dd}\\
&=& \frac{i (-1)^n}{(4 \pi)^{D/2}} \frac{1}{(8\pi)^{2-n}} \frac{\Gamma(n-D/2)}{\Gamma(n)} \left(\delta_{[D]0} + \delta_{[D]2}\right) \Lambda_\wp^{4-2n} \left(\frac{\mu}{m}\right)^{2n-D} \label{int-Dd-2}\\
&+& \frac{i (-1)^n}{(4 \pi)^{D/2}} \frac{\Gamma(n-D/2)}{\Gamma(n)} \delta_{[D]4} \; \mu^{4-2n} \left(\frac{\mu}{m}\right)^{2n-D} \label{int-Dd-3}
\end{eqnarray}
in which $[D]$ designates the integer part of $D$ so that $[0-\epsilon]=0, [2-\epsilon]=2$ and $[4-\epsilon]=4$ for an infinitesimal $\epsilon$. Needless to say, $\delta_{ij}$ is Kronecker delta, which is equal to 1 (0) if $i=j$ ($i\neq j$).  The normalization factor $1/(8\pi)^{2-n}$ is attached to make  coefficients of the $\Lambda_\wp^{4}$ and $\Lambda_\wp^{2}$ to remain parallel, respectively, to those of the quartic and quadratic terms cutoff regularization \cite{cutoff1}.

The loop integral (\ref{int-Dd-2}) is power-law in $\Lambda_\wp$ and logarithmic in $\mu$. The loop integral (\ref{int-Dd-3}), on the other hand, is independent of $\Lambda_\wp$ and logarithmic in $\mu$.  The first reason for this is that  dimensional regularization in $D=0$ and $D=2$ dimensions gives the quartic and quadratic UV divergences, respectively  \cite{jj1,jj2,jj3,jj4}. The second reason is that the power of $\Lambda_\wp$ is independent of $D$ but that of $\mu$ depends on $D$ and it gives rise therefore to  $\log(\mu/m)$ terms when the gamma functions are expanded about momentum space dimensions $D=0,2,4$. It is in this sense that the regularization method in (\ref{int-Dd}) leads to detached regularization simply because while $\Lambda_\wp$ arises only in power-law terms  $\mu$ appears only in logarithmic terms, and hence, the power-law and logarithmic UV-sensitivities get completely detached. This detachment can be explicitly seen by evaluating $I_{n,D}(\Lambda_\wp,\mu)$ for the relevant values of $D$ and $n\leq D/2$
\begin{eqnarray}
I_{n,D}(\Lambda_\wp,\mu) = \left\{\begin{array}{ll} \frac{i}{32 \pi^2} \Lambda_\wp^4 & \quad n=0, D=0\\ \\
 - \frac{i}{32 \pi^2} \Lambda_\wp^2\log \frac{\mu^2}{m^2} & \quad n=1, D=2\\ \\
\frac{i m^2}{16 \pi^2} \left(1 + \log \frac{\mu^2}{m^2}\right) & \quad n=1, D=4\\ \\
\frac{i}{16 \pi^2} \log \frac{\mu^2}{m^2} & \quad n=2, D=4
\end{array}\right.
\label{int-Dd-results}
\end{eqnarray}
after employing the $\overline{MS}$ subtraction scheme  \cite{gauge-break2, MSbar-2,MSbar-1}.  It is not surprising that the last two lines here are the same as the dimensional regularization results in (\ref{DR-2}). In general, detached regularization  provides a framework in which gauge breaking in the Wilsonian renormalization can be isolated and treated independently of the remaining dimensional regularization type corrections  \cite{bizimki,italyan}. 

\subsection{Gauge Symmetry Restoration and Emergent Gravity}
\label{Effective QFT}
The effective action capturing  physics of the full QFT at low energies can be structured as \cite{bizimki}
\begin{eqnarray}
S_{eff}[\eta, \digamma; \Lambda_\wp^2, \log\mu] = S_{tree}[\eta, \digamma] +  \delta S_{log}[\eta, \log\mu, \digamma] + \delta S_{pow}[\eta, \digamma; \Lambda_\wp^2, \log\mu]
\label{eff-ac}
\end{eqnarray}
under the detached regularization scheme in (\ref{int-Dd}). The first term $S_{tree}[\eta, \digamma]$ is the tree-level action. It sets  symmetries, field spectrum and interactions in the QFT.  The second term $\delta S_{log}[\eta, \log\mu, \digamma]$ collects the logarithmic loop corrections like correction actions (\ref{eff-ac-DR}) and (\ref{eff-ac-DR-2}) in the previous section. It maintains symmetries and spectrum of $S_{tree}[\eta, \digamma]$ and involves only the renormalization scale $\mu$ (not the UV cutoff $\Lambda_\wp$). The third term 
\begin{eqnarray}
\delta S_{pow}[\eta, \digamma; \Lambda_\wp^2, \log\mu]  &=& \int d^4x \sqrt{-\eta} \left\{-c_O \Lambda_\wp^4 - {\mathscr{M}^2} \Lambda_\wp^2 - c_\phi \Lambda_\wp^2 \phi^\dagger \phi + c_V \Lambda_\wp^2 {\rm tr}\left[V_\mu V^\mu\right] \right\} 
\label{Sp}
\end{eqnarray}
is the power-law correction obtained via the loop integrals in (\ref{int-Dd-results}). The loop factors characterizing it are given by \cite{bizimki}
\begin{eqnarray}
c_O&=&c_O(\log\mu)\xrightarrow{\rm one\ loop} \frac{(n_b-n_f)}{64 \pi^2},\label{cO-ilk}\\ {\mathscr{M}^2} &=& {\mathscr{M}^2}(\log\mu) \xrightarrow{\rm one\ loop} -\frac{1}{64\pi^2} {\rm str}\left[M^2 \log \frac{M^2}{\mu^2}\right],\label{MPl-ilk}\\ 
c_\phi&=&c_\phi(\log\mu)\xrightarrow{\rm one\ loop} \frac{\lambda_\phi}{16\pi^2},\label{cphi-ilk}\\   
c_V&=&c_V(\log\mu)\xrightarrow{\rm one\ loop} -\frac{g_V^2}{16\pi^2} C_r (n_b-n_f)^{(r)}
\label{cV-ilk}
\end{eqnarray}
in which $n_b(n_f)$ is the total number of bosons (fermions), $C_r$ and $(n_b-n_f)^{(r)}$ are, respectively, the quadratic Casimir and boson-fermion number difference in the representation $r$ of the gauge group, $M^2$ is the mass-squared matrix of the QFT fields, and $\lambda_{\phi}$ is a model-dependent coupling constant. At one loop, $c_{O} \propto n_b-n_f$ and this means that in QFTs with  boson-fermion balance (like supersymmetry) quartic contribution to vacuum energy vanishes identically. All the couplings $c_O,\dots, c_V$ have been calculated explicitly in \cite{bizimki} for both the scalar and spinor electrodynamics. 

\subsubsection{Promoting Poincare-Breaking UV Cutoff to Affine Curvature}
\label{Gauge Symmetry Restoration at the Extremum of Action}
The loop-induced gauge boson mass $c_V \Lambda_\wp^2$ in (\ref{Sp}) breaks the gauge symmetry. It breaks also the Poincare (translation) symmetry since  $\Lambda_\wp$ itself breaks the Poincare symmetry. The goal is to defuse this anomalous mass to restore the gauge symmetries. The Poincare-conserving vector boson mass $M_\wp$, as was already investigated in Sec. \ref{Gauge Symmetry Restoration at the Extremum of Energy},  can be promoted as in (\ref{MV-map}) to a Higgs field $\Phi$  to restore the gauge symmetries.   For a Poincare-breaking gauge boson mass $c_V \Lambda_\wp^2$, however, scalar fields like $\Phi$ cannot serve the purpose \cite{review1,review2,review3,review4}. The reason is that they are Poincare-conserving fields. It is necessary to find a ``Poincare-breaking field" such that  the gauge- and Poincare-breaking gauge boson mass $c_V \Lambda_\wp^2$ can be promoted to that particular field to restore the gauge symmetries. On physical grounds, the sought-for Poincare-breaking field is expected to be the spacetime curvature itself. In fact, in a general second-quantized field theory with no presumed symmetries,  Poincare (translation) invariance is known to emerge if the Poincare-breaking terms are identified with the spacetime curvature \cite{fn2}. It sounds consistent but there is a problem here.  The problem is that the Poincare-breaking gauge boson mass $c_V \Lambda_\wp^2$  arises in the flat spacetime effective QFT and it must therefore be promoted to curvature already in the flat spacetime. (QFTs are intrinsic to flat spacetime \cite{wald1,wald2}.) This means that the flat spacetime must somehow be endowed with the notion of curvature. This curvature cannot certainly be a metrical curvature since that would identically vanish in the flat metric limit. It can nevertheless be a metric-independent affine curvature as that would not vanish in the flat spacetime \cite{schroedinger-conn}.  Under a general coordinate transformation $x^\lambda \rightarrow (x^\prime)^\lambda \equiv x^{\lambda^\prime}$, any connection $\Gamma^\lambda_{\mu\nu}$ (metrical or affine) transforms as \cite{mtw-conn}
\begin{eqnarray}
  \Gamma^{\lambda^\prime}_{\mu^\prime\nu^\prime} = 
\frac{\partial x^{\lambda^\prime}}{\partial x^\lambda} \frac{\partial x^{\mu}}{\partial x^{\mu^\prime}} \frac{\partial x^{\nu}}{\partial x^{\nu^\prime}}   
  \Gamma^\lambda_{\mu\nu} + \frac{\partial x^{\lambda^\prime}}{\partial x^\lambda} \frac{\partial^2 x^{\lambda}}{\partial x^{\mu^\prime} \partial x^{\nu^\prime}}
  \label{conn-affine}
\end{eqnarray}
and while it transforms like this its Ricci curvature \cite{affine1,affine2,affine3}
\begin{eqnarray}
{\mathbb{R}}_{\mu\nu}(\Gamma) = \partial_\lambda \Gamma^\lambda_{\mu\nu} -\partial_\nu \Gamma^\lambda_{\mu\lambda} + \Gamma^\rho_{\lambda\rho} \Gamma^\lambda_{\mu\nu}  - \Gamma^\lambda_{\rho\nu}\Gamma^\rho_{\mu\lambda}
\label{affine-curv}
\end{eqnarray}
transforms as a tensor field
${\mathbb{R}}_{\mu^\prime\nu^\prime}(\Gamma^{\prime})=  \frac{\partial x^{\mu}}{\partial x^{\mu^\prime}} \frac{\partial x^{\nu}}{\partial x^{\nu^\prime}} {\mathbb{R}}_{\mu\nu}(\Gamma)$. Flat spacetime is characterized by the linear coordinate transformations $x^{\lambda^\prime}=a^{\lambda^\prime}_\lambda x^\lambda + b^{\lambda^\prime}$ and under these transformations $\frac{\partial^2 x^{\lambda}}{\partial x^{\mu^\prime} \partial x^{\nu^\prime}}=0$ in (\ref{conn-affine}) and, as a result, $\Gamma^\lambda_{\mu\nu}$  turns to a tensor field   ${\overline{\Gamma}}^\lambda_{\mu\nu}$ transforming as ${\overline{\Gamma}}^{\lambda^\prime}_{\mu^\prime\nu^\prime} = a^{\lambda^\prime}_\lambda a^{\mu}_{\mu^\prime} a^{\nu}_{\nu^\prime}
  {\overline{\Gamma}}^\lambda_{\mu\nu}$. If $\Gamma^\lambda_{\mu\nu}$  is a metrical connection then its flat spacetime value ${\overline{\Gamma}}_{\lambda\mu\nu}$ vanishes identically: $2{\overline{\Gamma}}_{\lambda\mu\nu} = \partial_\mu \eta_{\nu\lambda}+ \partial_\nu \eta_{\lambda\mu}-\partial_\lambda \eta_{\mu\nu} \equiv 0$.  If $\Gamma^\lambda_{\mu\nu}$  is a non-metrical affine connection then there is clearly no reason for its flat spacetime value ${\overline{\Gamma}}^\lambda_{\mu\nu}$ to vanish. And, as a result, 
  there remains a nonzero  affine  curvature ${\mathbb{R}}_{\mu\nu}({\overline{\Gamma}})$ in flat spacetime. This flat spacetime affine curvature, obtained from (\ref{affine-curv}) for $\Gamma^\lambda_{\mu\nu}={\overline{\Gamma}}^\lambda_{\mu\nu}$, enables  the Poincare-breaking gauge boson mass term in (\ref{Sp}) to be transformed into an affine curvature-gauge field interaction 
  \begin{eqnarray}
\Lambda_\wp^2 {\rm tr}\left[V_\mu \eta^{\mu\nu} V_\nu\right]
\longmapsto {\rm tr}\left[V_\mu {\mathbb{R}}^{\mu\nu}({\overline{\Gamma}}) V_\nu\right]
\label{MV-map-new}
\end{eqnarray}
in close similarity to the transformation in (\ref{MV-map}) of the Poincare-conserving gauge boson mass term to a scalar-vector interaction. This transformation promotes the UV cutoff $\Lambda_\wp$ to affine curvature as  $\Lambda_\wp^2 \eta_{\mu\nu} \longmapsto {\mathbb{R}}^{\mu\nu}({\overline{\Gamma}})$, and this way takes the power-law correction action $\delta S_{pow}[\eta, \digamma; \Lambda_\wp^2, \log\mu]$ in  (\ref{Sp}) to a new form
\begin{eqnarray}
\delta S_{pow}[\eta, \digamma; {\mathbb{R}}, \log\mu]  &=& \int d^4x \sqrt{-\eta}\, \left\{- \frac{c_O}{16} \left({\mathbb{R}}(\eta,{\overline{\Gamma}})\right)^2 
-\frac{{\mathcal{M}}^2}{4}{\mathbb{R}}(\eta,{\overline{\Gamma}})   -\frac{c_\phi}{4} \phi^\dagger\phi\, {\mathbb{R}}(\eta,{\overline{\Gamma}})  + c_V {\rm tr}\left[V_{\mu}{\mathbb{R}}^{\mu\nu}({\overline{\Gamma}})V_\nu \right] \right\}
\label{Sp-curv0}
\end{eqnarray}
in which ${\mathbb{R}}(\eta,{\overline{\Gamma}})\equiv \eta^{\mu\nu}{\mathbb{R}}_{\mu\nu}({\overline{\Gamma}})$ is the scalar affine curvature in flat spacetime. This action becomes extremum under variations in
${\overline{\Gamma}}^\lambda_{\mu\nu}$ provided that $\delta (\delta S_{pow})/\delta {\overline{\Gamma}} = 0$, and this motion equation  leads to a determination of   ${\mathbb{R}}^{\mu\nu}({\overline{\Gamma}})$ as a function of the scalars $\phi$ and gauge fields $V_\mu$. This functional relation eliminates ${\overline{\Gamma}}^\lambda_{\mu\nu}$ and leaves behind an effective field theory involving only the QFT fields. This is not a UV completion. Besides, gauge symmetries remain broken. In essence,  ${\overline{\Gamma}}^\lambda_{\mu\nu}$ is  a spurion at a similar level as the scalar $\Phi$ in (\ref{MV-map}), and it needs be given appropriate dynamical content to make room for UV completion and gauge symmetry restoration.

\subsubsection{Taking Effective QFT to Curved Spacetime}
\label{Effective QFT in Curved Spacetime}


One way to give dynamical content to ${\overline{\Gamma}}^\lambda_{\mu\nu}$ is to depart from flat spacetime. An indication of this is the transformation property of the affine connection in (\ref{conn-affine}) according to which $\frac{\partial^2 x^{\lambda}}{\partial x^{\mu^\prime} \partial x^{\nu^\prime}}\neq 0$ leads inherently to curved spacetime.  Another indication is the spurion action (\ref{Sp-curv0}) whose extremization against variations in ${\Gamma}^\lambda_{\mu\nu}$  would lead to a determination of ${\mathbb{R}}^{\mu\nu}(\Gamma)$  as a function not only of the scalars $\phi$ and gauge fields $V_\mu$ but also of the metrical curvature if the spacetime is curved. Under these indications, it becomes a necessity to carry  the flat spacetime effective QFT 
\begin{eqnarray}
S_{eff}[\eta, \digamma; {\mathbb{R}}, \log\mu] = S_{tree}[\eta, \digamma] +  \delta S_{log}[\eta, \log\mu, \digamma] + \delta S_{pow}[\eta, \digamma; {\mathbb{R}}, \log\mu]
\label{eff-ac-2}
\end{eqnarray}
into spacetime of a curved metric $g_{\mu\nu}$. This carriage is realized by general covariance through the map \cite{equiv,covariance} 
\begin{eqnarray}
   \eta_{\mu\nu} \rightarrow g_{\mu\nu}\,,\;\; \partial_\mu \rightarrow \nabla_\mu  
   \label{covariance}
\end{eqnarray}
in which the covariant derivative $\nabla_\mu$ is that of the Levi-Civita connection 
\begin{eqnarray}
    {}^g\Gamma^\lambda_{\mu\nu} = \frac{1}{2} g^{\lambda\rho} \left( \partial_\mu g_{\nu\rho} + \partial_\nu g_{\rho\mu} - \partial_\rho g_{\mu\nu}\right)
    \label{LC}
\end{eqnarray}
so that $\nabla_\alpha g_{\mu\nu}=0$. But, for the metric $g_{\mu\nu}$ to be curved, the effective QFT in curved spacetime must involve curvature of $g_{\mu\nu}$ like, for example, the metrical Ricci curvature $R_{\mu\nu}({}^g\Gamma)$. In parallel with additions of the kinetic and potential energies for the scalar $\Phi$ in the action (\ref{PV-11}), one may consider adding  curvature terms to the action (\ref{eff-ac-2}) like, for example, the Einstein-Hilbert term $M_0^2 g^{\mu\nu} R_{\mu\nu}({}^g\Gamma)$. This, however, does not work. The reason is that in Sec. \ref{Gauge Symmetry Restoration at the Extremum of Energy} the Poincare-conserving vector boson mass $M_\wp$ and the added scalar $\Phi$ were all at the tree-level. In the action (\ref{Sp-curv0}), however, all interactions involve  loop-induced parameters and addition of curvature terms  becomes simply inconsistent since new parameters like $M_0$ act as bare quantities. In fact, such added bare terms would mean that the curvature sector of the original QFT was left unrenormalized  while the QFT sector was being renormalized. The same is valid for curvature terms constructed with the bare  parameters in $S_{tree}[\eta, \digamma]$. It therefore is clear that curvature terms must arise multiplied with appropriate loop factors $c_i$. It is also clear that  curvature can arise only in the gauge sector because definition of the Ricci curvature $[\nabla_\lambda,\nabla_\nu] V^\lambda = R_{\mu\nu}({}^g\Gamma) V^\mu$ involves the gauge fields $V_\mu$. In this regard,  gauge kinetic invariant ${\rm tr}\left[V_{\mu\nu} V^{\mu\nu}\right]$ turns out to be a potential source  of the metrical curvature $R_{\mu\nu}({}^g\Gamma)$. 

One way to make curvature arise is to utilize the bulk and boundary properties of the gauge kinetic invariant. Under the general covariance map in (\ref{covariance}),  gauge field strength remains unchanged $D_\mu V_\nu-D_\nu V_\mu = V_{\mu\nu}= {\mathcal{D}}_\mu V_\nu-{\mathcal{D}}_\nu V_\mu$,  with the gauge-covariant derivative $D_\mu=\partial_\mu+iV_\mu$ in flat spacetime and ${\mathcal{D}}_\mu=\nabla_\mu+iV_\mu$ in curved spacetime. This metric-independence enables curvature to arise once the bulk and boundary parts of the gauge kinetic invariant are isolated. For instance, in flat spacetime,  $V_{\mu\nu} V^{\mu\nu}= V^{\mu} \left(-D^2\eta_{\mu\nu} + D_{\mu}D_\nu + i V_{\mu\nu}\right) V^{\nu}+\partial_{\mu}(V_{\nu}V^{\mu\nu})$ where $\partial_{\mu}(V_{\nu}V^{\mu\nu})$ is the aforementioned boundary term. This equality ensures that the integrated gauge kinetic invariant \cite{demir3,demir2,demir1}
\begin{eqnarray}
I_V[\eta] &=& \int d^4x \sqrt{-\eta}  \frac{1}{2} {\rm tr}\!\left[V_{\mu\nu} V^{\mu\nu}\right]\label{IV}
\end{eqnarray}
is equal to 
\begin{eqnarray}
{\bar{I}}_V[\eta] &=&\int d^4x \sqrt{-\eta}\, {\rm tr}\!\left[V^{\mu} (-D^2\eta_{\mu\nu} + D_{\mu}D_\nu + i V_{\mu\nu}) V^{\nu}+\partial_{\mu}(V_{\nu}V^{\mu\nu})\right]
\label{IVtil}
\end{eqnarray}
under by-parts integration. Needless to say, their difference vanishes identically
\begin{eqnarray}
\delta I_V[\eta] = -I_V[\eta] + {\bar{I}}_V[\eta] = 0
\label{delta-duz}
\end{eqnarray}
as a trivial identity. And, with this, the flat spacetime effective action  (\ref{eff-ac-2}) can be  formally put in a new form
\begin{eqnarray}
{\overline S}_{eff}[\eta, \digamma; {\mathbb{R}}, \log\mu] = S_{tree}[\eta, \digamma] +  \delta S_{log}[\eta, \log\mu, \digamma] + \delta S_{pow}[\eta, \digamma; {\mathbb{R}}, \log\mu] + c_V \delta I_V[\eta]
\label{eff-ac-3}
\end{eqnarray}
again as a trivial identity since addition of $c_V \delta I_V[\eta]$ changes nothing: ${\overline S}_{eff}[\eta, \digamma; {\mathbb{R}}, \log\mu]=S_{eff}[\eta, \digamma; {\mathbb{R}}, \log\mu]$. In a sense, this new effective action is a trivial regularization of the effective action $S_{eff}[\eta, \digamma; {\mathbb{R}}, \log\mu]$. The reason for multiplication by $c_V$ is to make curvature arise eventually in the form of a gauge boson mass term. In fact,  under the general covariance map in (\ref{covariance}), the effective action ${\overline S}_{eff}[\eta, \digamma; {\mathbb{R}}, \log\mu]$ above takes the form
\begin{eqnarray}
{\overline S}_{eff}[g, \digamma; {\mathbb{R}}, \log\mu] = S_{tree}[g, \digamma] +  \delta S_{log}[g, \log\mu, \digamma] + \delta S_{pow}[g, \digamma; {\mathbb{R}}, \log\mu] + c_V \delta I_V[g]
\label{eff-ac-4}
\end{eqnarray}
in which $\delta I_V[g]$, obtained from (\ref{IV}) and (\ref{IVtil}) under the general covariance map in (\ref{covariance}), takes a curvature-dependent nonzero value
\begin{eqnarray}
\delta I_V[g] = -I_V[g] + {\bar{I}}_V[g] = -\int d^4x \sqrt{-g}\, {\rm tr}\!\left[V^{\mu} R_{\mu\nu}({}^g\Gamma)V^{\nu}\right]
\label{delta-egri}
\end{eqnarray}
in manifest contrast to (\ref{delta-duz}). The reason for this contrast is that the gauge kinetic invariant in curved spacetime
$V_{\mu\nu} V^{\mu\nu} = V^{\mu} \left(-{\mathcal{D}}^2 g_{\mu\nu} + {\mathcal{D}}_{\mu}{\mathcal{D}}_\nu + i V_{\mu\nu} + R_{\mu\nu}({}^g\Gamma)) V^{\nu}+\nabla_{\mu}(V_{\nu}V^{\mu\nu}\right)$ differs from  the gauge kinetic invariant in flat spacetime $V_{\mu\nu} V^{\mu\nu}= V^{\mu} \left(-D^2\eta_{\mu\nu} + D_{\mu}D_\nu + i V_{\mu\nu}\right) V^{\nu}+\partial_{\mu}(V_{\nu}V^{\mu\nu})$   not only by the covariant derivative ${\nabla}_{\mu}$ but also by the Ricci curvature $R_{\mu\nu}({}^g\Gamma)$ generated by the commutator $[{\mathcal{D}}_\lambda,{\mathcal{D}}_\nu] V^\lambda = (R_{\mu\nu}({}^g\Gamma) + i V_{\mu\nu}) V^\mu$.
This difference between the invariants leads to the relation $I_V[g]={\bar{I}}_V[g]+\int d^4x \sqrt{-g}\, {\rm tr}\!\left[V^{\mu} R_{\mu\nu}({}^g\Gamma)V^{\nu}\right]$, and from this  follows the $\delta I_V[g]$ in (\ref{delta-egri}). Effectively, as far as the gauge boson mass term is concerned, transfer of the flat spacetime effective action in (\ref{eff-ac-2}) to curved spacetime implies the map
\begin{eqnarray}
{\rm tr}\left[V_\mu {\mathbb{R}}^{\mu\nu}({\overline{\Gamma}}) V_\nu\right]
\longmapsto {\rm tr}\left[V_\mu \left({\mathbb{R}}^{\mu\nu}(\Gamma)- R^{\mu\nu}({}^g\Gamma)\right)V_\nu\right]
\label{MV-map-new-curv}
\end{eqnarray}
as succession of the promotion in (\ref{MV-map-new}) of the Poincare-breaking UV cutoff $\Lambda_\wp$ to flat spacetime affine curvature ${\mathbb{R}}({\overline{\Gamma}})$, where different connections and curvatures used in this map and earlier definitions are contrasted in Table \ref{table-1} for clarity.
\begin{table}
\caption{Contrasting the metrical and affine quantities in the flat and curved spacetimes. \label{table-1}}
	\begin{tabular}{l|l|l|l|l|l}
			\hline\hline
	{}&metric tensor&metrical connection & metrical curvature & affine connection & affine curvature\\ \hline 
 flat spacetime & $\eta_{\mu\nu}$& 0 & 0 &  ${\overline{\Gamma}}^{\lambda}_{\mu\nu}$ & ${\mathbb{R}}_{\mu\nu}({\overline{\Gamma}})$\\ \hline
curved spacetime & $g_{\mu\nu}$& ${}^g\Gamma^\lambda_{\mu\nu}$ & $R_{\mu\nu}({}^g\Gamma)$  &  $\Gamma^{\lambda}_{\mu\nu}$ & ${\mathbb{R}}_{\mu\nu}(\Gamma)$
\end{tabular}
\end{table}

In regard to the new map in (\ref{MV-map-new-curv}), the curved spacetime effective action in (\ref{eff-ac-4}) can be recast in the same form as  (\ref{eff-ac-2}) 
\begin{eqnarray}
{\overline S}_{eff}[g, \digamma; {\mathbb{R}}, \log\mu] = S_{tree}[g, \digamma] +  \delta S_{log}[g, \log\mu, \digamma] + \delta {\overline S}_{pow}[\eta, \digamma; {\mathbb{R}}, \log\mu]
\label{eff-ac-curv}
\end{eqnarray}
such that $\delta {\overline S}_{pow}[\eta, \digamma; {\mathbb{R}}, \log\mu] = \delta S_{pow}[g, \digamma; {\mathbb{R}}, \log\mu] + c_V \delta I_V[g]$ has the explicit form
\begin{eqnarray}
\delta {\overline{S}}_{pow}[g, \digamma; {\mathbb{R}}, \log\mu,R]  &=& \int d^4x \sqrt{-g} \Bigg\{\!\!
-\frac{M_{Pl}^2}{2}{\mathbb{R}}(g,\Gamma)   - \frac{{\overline{c}}_O}{16} \left({\mathbb{R}}(g,\Gamma)\right)^2  -\frac{c_\phi}{4} \phi^\dagger\phi\,  {\mathbb{R}}(g,\Gamma) \nonumber\\&&\qquad\qquad\quad\;\;+\, c_V {\rm tr}\left[V^{\mu}\!\left({\mathbb{R}}_{\mu\nu}(\Gamma)- R_{\mu\nu}({}^g\Gamma)\right)\!V^{\nu}\right]\!\Bigg\}
\label{Sp-curvp}
\end{eqnarray}
in which ${\mathbb{R}}(g,\Gamma)\equiv g^{\mu\nu}{\mathbb{R}}_{\mu\nu}(\Gamma)$ is the curved spacetime analog of the flat spacetime affine curvature ${\mathbb{R}}(\eta,{\overline{\Gamma}})$ in (\ref{Sp-curv0}).  The parameters in this action deserve a detailed discussion: First, as follows from the equation (\ref{MPl-ilk}), 
the fundamental scale of gravity (the Planck scale) can be defined as
\begin{eqnarray}
M_{Pl}^2= \frac{1}{2} {\mathscr{M}^2}\!\left(\mu=M_{Pl}\right) \xrightarrow{\rm one\ loop}  -\frac{1}{128\pi^2} {\rm str}\!\left[{\overline{M}}^2 \log \left(\frac{{\overline{M}}^2}{M_{Pl}^2}\right)\right]
\label{MPl}
\end{eqnarray}
in which ${\overline{M}}^2 =  M^2\big(\mu=M_{Pl}\big)$. This definition comes to mean that the gravitational scale is equal to ${\mathscr{M}^2}/2$ evaluated at the gravitational scale. It is clear that  $M_{Pl}$ remains put at its value in (\ref{MPl}) since curvature is classical and matter loops have already been used up in forming the flat spacetime effective action in (\ref{eff-ac}). It is also clear that gravity remains attractive if the bosonic sector is heavier (${\rm str}\!\big[{\overline{M}}^2\big]> 0$) and if all the matter fields weigh below the gravitational scale (${\rm str}\!\big[{\overline{M}}^2\big]\lesssim M_{Pl}^2$). In general,  scalars, singlet fermions and vector-like fermions can weigh heavy without breaking gauge symmetries, and such heavy fields can dominate $M_{Pl}$ through ${\rm str}\!\big[{\overline{M}}^2\big]$.

In the same sense as the gravitational scale in (\ref{MPl}),  quadratic curvature coefficient ${\overline{c}}_O$ in (\ref{Sp-curvp}) can be defined as
\begin{eqnarray}
{\overline{c}}_O=c_O\left(\mu=M_{Pl}\right)\xrightarrow{\rm one\ loop} \frac{(n_b-n_f)}{64 \pi^2} 
\label{cO}
\end{eqnarray}
and this definition means that $n_b (n_f)$ is the number of bosons (fermions) having masses from zero way up to the gravitational scale. In other words, $n_b (n_f)$  comprises entirety of the bosons (fermions) since particles heavier than $M_{Pl}$ are disfavored by the attractive nature of gravity. It is clear that  ${\overline{c}}_O$ remains set at its value in (\ref{cO}). 

In similarity to $M_{Pl}$ in (\ref{MPl}) and ${\overline{c}}_O$ in (\ref{cO}), the loop factors $c_\phi$ and $c_V$ in (\ref{Sp-curvp}) are set to their values at $\mu=M_{Pl}$. But, unlike the $M_{Pl}$ and ${\overline{c}}_O$, these two loop factors run with the renormalization scale $\mu$ since both the scalars $\phi$ and gauge bosons $V_\mu$ vary with the scale as quantum fields (wavefunction renormalization).  And their variations are governed by the flat spacetime matter loops as in (\ref{cphi-ilk}) and (\ref{cV-ilk}).

\subsubsection{Affine Dynamics: Gauge Symmetry Restoration and Emergent Gravity}
\label{Affine Dynamics: From UV Cutoff to IR Curvature}
The effective action \eqref{eff-ac-curv} remains stationary against variations in the affine connection ($\delta S_{eff}/\delta\Gamma = 0$) if 
\begin{eqnarray}
\label{gamma-eom}
{}^{\Gamma}\nabla_{\lambda}{{\mathbb{D}}}_{\mu\nu} = 0
\end{eqnarray}
such that ${}^{\Gamma}\nabla_{\lambda}$ is covariant derivative with respect to the affine connection $\Gamma^\lambda_{\mu\nu}$, ${\mathbb{D}}_{\mu\nu}= (g/{\rm Det}[{\mathbb{Q}}])^{1/6} {\mathbb{Q}}_{\mu\nu}$, and 
\begin{eqnarray}
\label{q-tensor}
{{\mathbb{Q}}}_{\mu\nu} = \left(\frac{1}{16\pi G_N} +   \frac{c_\phi}{4} \phi^\dagger\phi + \frac{{\overline{c}}_O}{8} g^{\alpha\beta} {\mathbb{R}}_{\alpha\beta}(\Gamma)\right) g_{\mu\nu} - c_{V} {\mbox{tr}}\left[V_{\mu}V_{\nu}\right]
\end{eqnarray}
is the disformal metric of tensor fields, including the affine curvature ${\mathbb{R}}(\Gamma)$ itself. 
The motion equation (\ref{gamma-eom}) implies that ${{\mathbb{D}}}_{\mu\nu}$ is covariantly-constant with respect to $\Gamma^\lambda_{\mu\nu}$, and this constancy leads to the exact  solution
\begin{eqnarray}
\Gamma^\lambda_{\mu\nu} &=& \frac{1}{2} \left({{ \mathbb{D}}}^{-1}\right)^{\lambda\rho} \left( \partial_\mu {{ \mathbb{D}}}_{\nu\rho} + \partial_\nu {{ \mathbb{D}}}_{\rho\mu} - \partial_\rho {{ \mathbb{D}}}_{\mu\nu}\right)\nonumber\\ &=& {}^g\Gamma^\lambda_{\mu\nu} + \frac{1}{2} ({{\mathbb{D}}}^{-1})^{\lambda\rho} \left( \nabla_\mu {{ \mathbb{D}}}_{\nu\rho} + \nabla_\nu {{ \mathbb{D}}}_{\rho\mu} - \nabla_\rho {{\mathbb{D}}}_{\mu\nu}\right)
\label{aC}
\end{eqnarray}
where the Levi-Civita connection ${}^g\Gamma^\lambda_{\mu\nu}$ was defined in (\ref{LC}). The Planck scale in (\ref{MPl})  is the largest scale and therefore it is legitimate to make the expansions
\begin{eqnarray}
\Gamma^{\lambda}_{\mu\nu}&=&{}^{g}\Gamma^{\lambda}_{\mu\nu} + \frac{1}{M_{Pl}^2} \left( \nabla_\mu {\overline{\mathbb D}}^\lambda_\nu + \nabla_\nu {\overline{\mathbb D}}^\lambda_\mu - \nabla^\lambda {\overline{\mathbb D}}_{\mu\nu}\right) + {\mathcal{O}}\left(M_{Pl}^{-4}\right)
\label{expand-conn}
\end{eqnarray}
and
\begin{eqnarray}
{\mathbb{R}}_{\mu\nu}(\Gamma) &=& R_{\mu\nu}({}^{g}\Gamma) + \frac{1}{M_{Pl}^2}\left(\nabla^{\alpha} \nabla_{\mu} {\overline{\mathbb D}}_{\alpha\nu} + \nabla^{\alpha} \nabla_{\nu} {\overline{\mathbb D}}_{\alpha\mu} - \Box {\overline{\mathbb D}}_{\mu\nu} - \nabla_{\mu} \nabla_{\nu} {\overline{\mathbb D}}_{\alpha}^{\alpha}\right)  +  {\mathcal{O}}\left(M_{Pl}^{-4}\right)
\label{expand-curv}
\end{eqnarray}
so that both $\Gamma^{\lambda}_{\mu\nu}$ and ${\mathbb{R}}_{\mu\nu}(\Gamma)$ contain pure derivative terms  at  
the next-to-leading order through the reduced disformal metric  ${\overline{\mathbb D}}_{\mu\nu}=
\left(\frac{c_\phi}{12} \phi^\dagger\phi + \frac{{\overline{c}}_O}{24} g^{\alpha\beta} {\mathbb{R}}_{\alpha\beta}(\Gamma)+\frac{c_V}{6} g^{\alpha\beta} {\mbox{tr}}\left[V_{\alpha}V_{\beta}\right]\right) g_{\mu\nu} - c_{V} {\mbox{tr}}\left[V_{\mu}V_{\nu}\right]$ \cite{demir1,demir2,demir3,bks}. The expansion in (\ref{expand-conn}) ensures that the affine connection $\Gamma^{\lambda}_{\mu\nu}$ is solved algebraically order by order in $1/M_{Pl}^{2}$  despite the fact that its motion equation (\ref{gamma-eom}) involves its own curvature ${\mathbb{R}}_{\mu\nu}(\Gamma)$ through ${\mathbb{D}}_{\mu\nu}$   \cite{affine1,affine2}. The expansion (\ref{expand-curv}), on the other hand, ensures that the affine curvature  ${\mathbb{R}}_{\mu\nu}(\Gamma)$ is equal to the metrical curvature $R_{\mu\nu}({}^g\Gamma)$ up to a doubly-Planck suppressed remainder. In essence, what happened is that the affine dynamics took the affine curvature ${\mathbb{R}}$ from its UV value $\Lambda_\wp^2$  in (\ref{MV-map-new}) to its IR value $R$ in (\ref{expand-curv}). Indeed,  in the sense of holography \cite{holog0,holog1}, the metrical curvature $R$ sets the IR scale \cite{holog2} above which QFTs hold as flat spacetime constructs. 

\begin{enumerate}
\item {\bf Gauge Symmetry Restoration at the Extremum of Affine Action.} 
One consequence of the solution of the affine curvature in (\ref{expand-curv}) is that the problematic loop-induced gauge boson mass term gets defused as
\begin{eqnarray}
\int d^4x \sqrt{-g} c_V {\rm tr}\left[V^{\mu}\!\left({\mathbb{R}}_{\mu\nu}(\Gamma)- R_{\mu\nu}({}^g\Gamma)\right)\!V^{\nu}\right] \xrightarrow{\rm equation\, (\ref{expand-curv})} \int d^4x \sqrt{-g} \!\left\{ {\rm zero} + {\mathcal{O}}\!\left(M_{Pl}^{-2}\right)\!\right\}
\label{reduce-gauge} 
\end{eqnarray}
after using the solution of the affine curvature in (\ref{expand-curv}) in the metric-Palatini action (\ref{Sp-curvp}) \cite{demir1,demir2,demir3}. The ${\mathcal{O}}\!\left(M_{Pl}^{-2}\right)$ remainder here, containing the next-to-leading order derivative term in (\ref{expand-curv}),   involves derivatives  of the scalars $\phi$ and gauge fields $V_\mu$, and can  produce therefore  no mass terms for either of them. It is worth noting that the gauge symmetries broken explicitly by a Poincare-conserving (Poincare-breaking) UV cutoff are restored via the Higgs field $\Phi$  (via the affine curvature ${\mathbb{R}}$) at the minimum of the $\Phi$ potential energy (at the extremum of the metric-affine action). This contrast shows that Poincare-conserving and Poincare-breaking UV cutoffs are fundamentally different and lead, respectively, to field-theoretic and gravitational completions of the effective QFT.

\item {\bf Emergence of Gravity at the Extremum of Affine Action.} 
One other consequence of the solution of the affine curvature in (\ref{expand-curv}) is that non-gauge sector of the metric-Palatini action (\ref{Sp-curvp}) reduces to the quadratic curvature gravity
\begin{eqnarray}
     \int d^4x \sqrt{-g} \Bigg\{\!
&& - \frac{M_{Pl}^2}{2}{\mathbb{R}}(g,\Gamma)   - \frac{{\overline{c}}_O}{16} \left({\mathbb{R}}(g,\Gamma)\right)^2 -\frac{c_\phi}{4} \phi^\dagger\phi\,  {\mathbb{R}}(g,\Gamma)\! \Bigg\}\nonumber\\
&&\xrightarrow{\rm equation\, (\ref{expand-curv})}
\int d^4x \sqrt{-g} \Bigg\{\!
-\frac{M_{Pl}^2}{2}R   - \frac{{\overline{c}}_O}{16} R^2  -\frac{c_\phi}{4} \phi^\dagger\phi\, R + {\mathcal{O}}\!\left(M_{Pl}^{-2}\right)\!\Bigg\}
\label{reduce-nongauge}
\end{eqnarray}
in which $R=g^{\mu\nu}R_{\mu\nu}({}^g\Gamma)$ is the usual curvature scalar in the GR. As in (\ref{reduce-gauge}), the ${\mathcal{O}}\left(M_{Pl}^{-2}\right)$ remainder here consists of the next-to-leading order and higher terms  in (\ref{expand-curv}). It  involves derivatives  of the long-wavelength fields $\phi$ and $V_\mu$, produces thus  no mass terms for these fields, and remains small for all practical purposes.  

The reductions (\ref{reduce-gauge}) and (\ref{reduce-nongauge}) give rise to the total QFT plus GR action 
\begin{eqnarray}
S_{tot}[g, \digamma] &=& S_{tree}[g, \digamma] +  \delta S_{log}[g, \log\mu, \digamma] +
\!\!\int\! d^4x \sqrt{-g} \left\{\!
-\frac{M_{Pl}^2}{2}R -\frac{c_O}{16} R^2 - \frac{c_\phi}{4} \phi^\dagger\phi R\!\right\}
\label{eff-ac-curvpp}
\end{eqnarray}
in which the QFT sector 
\begin{eqnarray}
S_{\small {\rm QFT}}[g, \digamma] =  S_{tree}[g, \digamma] +  \delta S_{log}[g, \log\mu, \digamma]
\label{QFT-sector}
\end{eqnarray}
is the usual ${\overline{MS}}$-renormalized QFT resting on the  matter loops in flat spacetime and evolving from scale to scale by renormalization group equations in  $\log\mu$ \cite{gauge-break2,MSbar-1,MSbar-2}. Its gravity sector
\begin{eqnarray}
S_{{\rm GR}}[g, \phi] =\int d^4x \sqrt{-g} \left\{
-\frac{M_{Pl}^2}{2} R - \frac{c_O}{16} R^2 - \frac{c_\phi}{4} \phi^\dagger\phi R \right\}
\label{gravity-sector}
\end{eqnarray}
rests on the flat spacetime loop factors and emerges from the requirement of restoring gauge symmetries. With these two unique features, it differs  from all the other matter-induced gravity theories (induced \cite{induced-grav1,induced-grav2,induced-grav3}, emergent \cite{emerge-grav1,emerge-grav2}, analogue \cite{analogue-grav1}, broken symmetry \cite{brokensymm-grav1, brokensymm-grav2}, and the like). It is an $R+R\phi^2+R^2$ gravity theory \cite{R+Rkare,R+Rkare2} whose each and every coupling is a flat spacetime loop factor (coefficient of $\Lambda_\wp^2$ or $\Lambda_\wp^4$ in (\ref{Sp})).  It is  the gauge symmetry-restoring emergent gravity or briefly the {\it symmergent gravity} \cite{demir3,demir2,demir1}, which is reformulated in a completely new setting in the present work. It should not be confused with the effective action computed in curved spacetime, which gives $M_{Pl}^2\propto \Lambda_\wp^2$ along with $\Lambda_\wp$-sized scalar and gauge boson masses and a $\Lambda_\wp^4$-sized vacuum energy density \cite{induced-grav1,induced-grav2,induced-grav3}. Symmergent gravity, as reformulated and elucidated in the present work, stands out as a novel framework for completing effective QFTs in the UV when the UV cutoff is a Poincare-breaking one. 
\end{enumerate}

\subsection{Salient Implications of the Symmergent Gravity}
\label{Salient Properties of the Emergent Gravity}
Symmergent gravity has a number of physics implications. It is worth touching on some salient ones here:

\begin{enumerate}[(a)]
\item {\bf Gravitational scale necessitates new physics beyond SM (BSM).} 
The gravitational scale $M_{Pl}$ in $S_{\rm GR}(g,\phi)$ is induced by  the flat spacetime matter loops as in (\ref{MPl}).  In the SM, it takes the value $M_{Pl}^2 \approx - G_F^{-1}$, where $G_F\approx (293\ {\rm GeV})^{-2}$ is the Fermi scale. Its negative sign, set by the top quark contribution, is obviously unacceptable. It must be turned to positive if gravity is to be attractive, and this can be done if there exist new particles beyond the SM spectrum.  These beyond-the-SM (BSM) particles are a necessity.

Inclusion of the BSM fields has turned $M_{Pl}^2$ to positive. This, however, is not sufficient because fundamental scale of gravity  must take its physical value of ${\overline M}_{Pl}\approx 10^{18}\ {\rm GeV}$. And gravity can be given this physical fundamental scale by rescaling the metric as 
\begin{eqnarray}
g_{\mu\nu} \longrightarrow \frac{{\overline M}^2_{Pl}}{M_{Pl}^2}\, g_{\mu\nu}
\label{conf-trans}
\end{eqnarray}
since then $\sqrt{g} M_{Pl}^2 R  \rightarrow \sqrt{g} {\overline M}^2_{Pl} R $ \cite{conformal,conformal2}. This conformal transformation affects the entire matter plus gravity action (\ref{eff-ac-curvpp}) in accordance with the equivalence principle \cite{equiv}. And its effects on coupling constants in the action become stronger and stronger as  the ratio ${\overline M}^2_{Pl}/M_{Pl}^2$ gets larger and larger. In this sense, it is necessary to keep  ${\overline M}^2_{Pl}/M_{Pl}^2$ close to unity to maintain the loop-induced structure of the QFT and the emergent gravity. Incidentally, an enormous value like  $M_{Pl}^2 \simeq {\overline M}^2_{Pl}$ was the main assumption behind the restoration of gauge symmetries in (\ref{reduce-gauge}) and emergence of gravity in (\ref{reduce-nongauge}). Thus, it is consistent to continue with  $M_{Pl}^2 \simeq {\overline M}^2_{Pl}$ keeping in mind that a conformal transformation like (\ref{conf-trans}) can always set the correct gravitational scale ${\overline M}_{Pl}$. In this regard, the BSM sector must have
\begin{enumerate}[(i)]
    \item either a light spectrum with numerous more bosons than fermions (for instance, $m_b \sim m_f \sim G_F^{-1/2}$ with $n_b-n_f \sim 10^{32}$),
    
    \item or a heavy spectrum with few more bosons than fermions (for instance, $m_b \sim m_f \lesssim {\overline M}_{Pl}$ with $n_b-n_f \gtrsim 10$),
    
    \item or a sparse spectrum with net boson dominance,
\end{enumerate}
to satisfy the constraint $M_{Pl}^2 \simeq {\overline M}^2_{Pl}$. In general, BSM particles do not have to couple to the SM particles simply because all they are requiered to do is to saturate the super-trace in (\ref{MPl}) at a value $M_{Pl}^2 \simeq {\overline M}^2_{Pl}$. In other words, there  are no symmetries or selection rules requiring the SM particles to couple to the BSM particles. They can form therefore a fully-decoupled black
sector \cite{black-1,black-2,black-3,demir2} or a feebly-coupled dark sector \cite{demir2,darksector-weak,darksector-natural}, with distinctive
signatures at collider searches \cite{keremle}, dark matter searches  \cite{cemle}, and other  possible phenomena \cite{demir2}.

\item {\bf Higgs mechanism remains intact.} 
In constructing the symmergence, entire analysis has been restricted to massless gauge fields like the gluon. But, actually, the methodology summarized by equation (\ref{reduce-gauge}) applies equally well to massive vector fields like the W/Z bosons in the SM because all that matters are the power-law corrections in (\ref{Sp}). These massive vectors acquire their masses from the Higgs mechanism. And the Higgs potential receives loop corrections from  $\delta S_{log}[g, \log\mu, \digamma]$ in (\ref{QFT-sector}). These logarithmic corrections are identical to what one would find by using the dimensional regularization (thanks to the detached regularization). In symmergent gravity, therefore, the Higgs mechanism is expected to work as usual to generate masses for massive fields (modulo the corrections from the BSM particles inferred in part (a) above).  Needless to say, symmergence works in the same way for all the fields in the BSM sector, including the massless and massive vector fields therein.

\item {\bf Higgs-curvature coupling can probe the BSM.} The loop factor $c_\phi$ in  $S_{\rm GR}(g,\phi)$ couples the scalar curvature $R(g)$ to scalar fields $\phi$. It is about $1.3\%$ in the SM \cite{demir1,demir2}. Its deviation from this SM value indicates existence of  new particles which couple to the SM Higgs boson. These BSM particles  can be probed via their effects on various gravitational and astrophysical phenomena \cite{non-min-1,non-min-2,non-min-3}. 

\item {\bf BSM symmetries might shed new light on the cosmological constant problem.} The vacuum energy contained in $S_{\rm QFT}(g, \digamma)$ \cite{bizimki}
\begin{eqnarray}
V(\mu) \xrightarrow{\rm one\ loop} V\left(\langle\phi\rangle\right) + \frac{1}{32\pi^2}  {\rm str} \left[{\overline{M}}^4 \left(1-\frac{3}{2}\log \frac{{\overline{M}}^2}{\mu^2}\right)\right]
\label{vac-en}
\end{eqnarray}
gathers together field-independent $\log\mu$ corrections in (\ref{QFT-sector}) in the minimum $\phi=\langle \phi\rangle$ of the scalar potential $V(\phi)$. Its
empirical value is $V_{emp} = \left(2.57\times 10^{-3}\ {\rm eV}\right)^4$ \cite{ccp2}. The cosmological constant problem is to shoot this specific value with the prediction in (\ref{vac-en}), and such a shooting is tantalizingly difficult to achieve \cite{ccp1}. Nevertheless, as a way out possible only in  symmergence, it might be possible to achieve a resolution if the BSM fields (which do not have to couple to the SM particles) enjoy efficacious symmetries and selection rules \cite{darksector-natural}. For instance, a supersymmetry-like structure in the BSM sector would kill the vacuum energy though realizing a partly-supersymmetric QFT may require extra structures \cite{tony}.

\item {\bf Quadratic curvature term can probe the BSM.} The loop factor $c_O$ in $S_{\rm GR}(g,\phi)$ is proportional to the boson-fermion number difference. It vanishes identically in a QFT with equal bosonic and fermionic degrees of freedom (as in the supersymmetric theories \cite{dimopoulos,susy}) and, as a result, the gravitational sector in (\ref{gravity-sector}) reduces to the GR. This normally is not possible given that, under general covariance, all curvature invariants can contribute to the gravitational sector, and it simply is not possible to get the exact GR. But symmergence is able to generate the GR, exact GR, when the SM+BSM involves  equal bosonic and fermionic degrees of freedom \cite{demir2,demir3}. Conversely speaking, if experiments and observations reveal that the quadratic curvature term is absent (namely, $c_O=0$) then it is for sure that the underlying SM+BSM has equal numbers of bosons and fermions. 

The loop factor $c_O$, when nonzero, acts as a vestige of strong curvatures. Indeed, it can probe the BSM sector in terms of $n_b-n_f$ via strong-curvature effects. One such effect is the Starobinsky inflation, and seems to require $n_b-n_f \approx 10^{13}$ \cite{irfan}. In addition, symmergent gravity itself may realize a quantum cosmology phase \cite{irfan2}. One other effect concerns the black holes, which put limits \cite{bh1,bh2,bh3,bh4} on $c_O$ and the vacuum energy $V(\mu)$  through the EHT observations \cite{EHT}. Effects of $c_O$ on stellar structure and wormhole dynamics can also be significant.

\item {\bf Heavy BSM does not necessarily destabilize the light scalars.} The light scalar fields $\phi_L$ in $S_{\rm QFT}(g, \digamma)$ qualify unnatural due to their oversensitivity to heavy fields. Indeed, as already revealed in Sec. \ref{Gauge-Conserving UV Cutoff}, their masses $m_{\phi_L}$ get shifted by  
\begin{eqnarray}
\delta m_{\phi_L}^2 = {c}_{\phi_L} \lambda_{\phi_L \digamma_{\!\!H}} m_{\digamma_{\!\!H}}^2 \log \frac{m_{\digamma_{\!\!H}}^2}{\mu^2} 
\label{deltmphi-lhp}
\end{eqnarray}
if they couple with loop factor ${c}_{\phi_L}$ and coupling constant  $\lambda_{\phi_L \digamma_{\!\!H}}$ to heavy fields $\digamma_{\!\!H}$ of masses  $m_{\digamma_{\!H}}\gg m_{\phi_L}$. This loop correction reveals that 
heavier the $\digamma_{\!\!H}$ larger the shift in the $\phi_L$ mass and stronger the destabilization of the light scalar sector. This is the well-known little hierarchy problem \cite{lhp1,lhp2}. In fact, null LHC results \cite{ATLAS,CMS} have sidelined supersymmetry and other known completions on the basis of this problem namely on the basis of large corrections like (\ref{deltmphi-lhp}) to the Higgs boson mass.

Symmergence is fundamentally different from the other UV completions like supersymmetry. It has the potential to provide a way out from the little hierarchy problem because the SM-BSM couplings are not constrained at all. Indeed, 
its BSM sector is necessitated only for generating the gravitational scale namely  saturating the super-trace formula (\ref{MPl}), and hence,  there is no symmetry principle or selection rule requiring the SM and BSM fields to interact. Namely, the coupling $\lambda_{\phi_L \digamma_{\!\!H}}$ is not under any constraint since workings of symmergence do not depend on its strength. This means that symmergence allows $\delta m_{\phi_L}^2$ to be small enough because  $\lambda_{\phi_L \digamma_{\!\!H}}$ is allowed to be small enough. More precisely, it is possible ensure $\delta m_{\phi_L}^2 \ll m_{\phi_L}^2$  because $\lambda_{\phi_L \digamma_{\!\!H}}$ can always be taken in the range \cite{demir2,demir3}
\begin{eqnarray}
\label{seesawic}
\left|\lambda_{\phi_L \digamma_{\!\!H}}\right| \lesssim \lambda_{SM} \frac{m_{\phi_L}^2}{m_{\digamma_{\!\!H}}^2}  
\end{eqnarray}
where $\lambda_{SM} \sim {\mathcal{O}}(1)$ is a typical SM coupling. This ``small-coupling domain" is specific to symmergence.  It is the domain in which the SM and BSM are sufficiently decoupled and the  little hierarchy problem is naturally avoided  \cite{darksector-natural}.  In contrast to supersymmetry and other completions, therefore, symmergence allows the SM and BSM to be decoupled as in (\ref{seesawic}) so that the light scalars like the Higgs boson remain stable or, in different words, the scale separation between the SM and BSM is maintained at the loop level. As a concrete example, one comes to realize that symmergence requires neutrinos to be Dirac as because neutrino masses and Higgs stability bound (\ref{seesawic}) cannot be satisfied at the same time with Majorana neutrinos \cite{darksector-natural,demir2,demir3}. In summary, the BSM of symmergence does not necessarily destabilize light scalars like the Higgs boson \cite{demir2,demir3}. Its BSM particles can  form a fully-decoupled black sector \cite{black-1,black-2,black-3,demir2} or a feebly-coupled dark sector \cite{demir2,darksector-weak,darksector-natural}.  And, needless to say, this naturally-coupled symmergent BSM agrees with the bulk of collider \cite{ATLAS,CMS}, cosmological \cite{ccp2} and astrophysical \cite{darksector-weak,darksector-natural} data.  
\end{enumerate}

\section{Poincare-Conserving UV Cutoff vs. Poincare-Breaking UV Cutoff: Contrasting the Two UV Completions}
\label{Poincare-Conserving vs Poincare-Breaking UV Cutoffs}

\begin{figure}[ht!]
  \includegraphics[width=18cm]{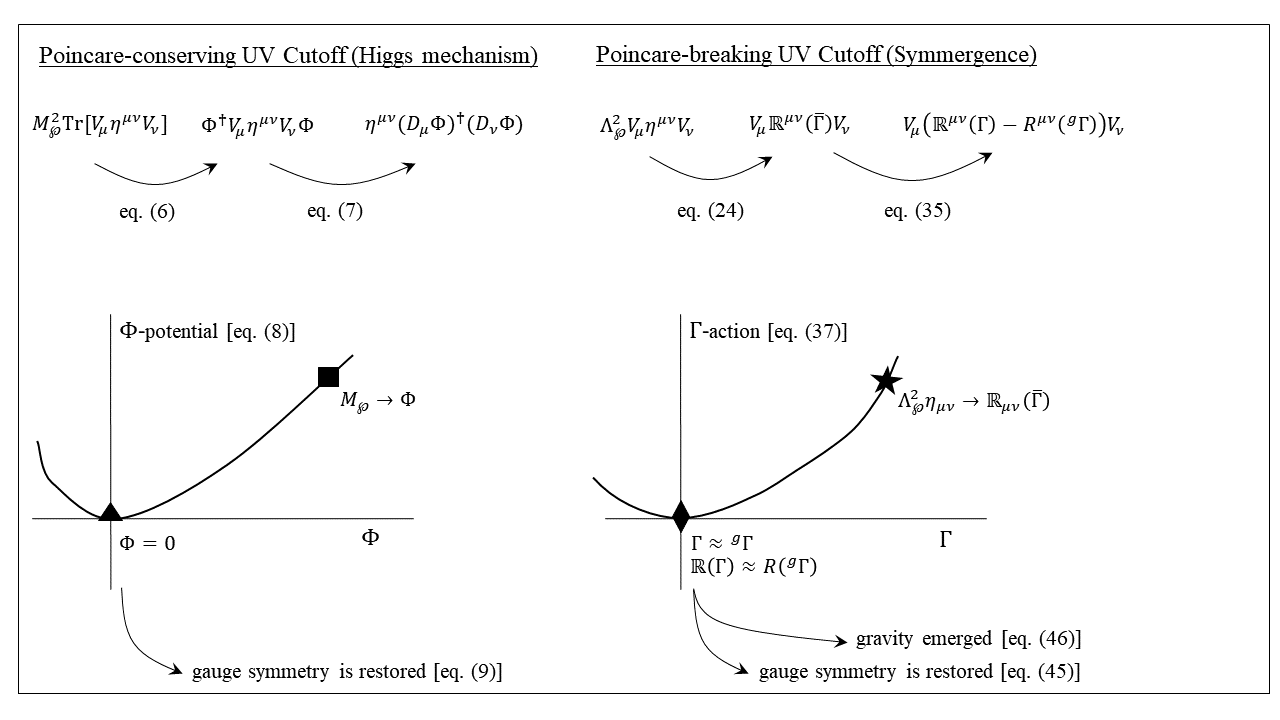} 
\caption{Gauge symmetry restoration and UV completion  in effective QFTs with Poincare-conserving (left) and Poincare-breaking (right) UV cutoffs.  The $\Phi$-potential on the left (minimized as in (\ref{poten-min})) and ${\Gamma}$-action on the right (extremized as in (\ref{reduce-nongauge})) are schematic drawings.}
\label{fig1}
\end{figure}

Having constructed the UV completions for Poincare-conserving  and  Poincare-breaking UV cutoffs each, contrasting them  proves useful for elucidating their working principles and physics implications. 
In Sec. \ref{Gauge-Breaking UV Cutoff}, it is shown that the effective QFTs with Poincare-conserving and gauge-breaking UV cutoffs could be completed in the UV by the Higgs mechanism. In Sec. \ref{Poincare-Breaking UV Cutoff}, on the other hand, it is shown that the effective QFTs with Poincare- and gauge-breaking UV cutoffs could be completed in the UV by the symmergence mechanism. The two mechanisms have similarities and dissimilarities, and a stimulating way of contrasting them is the infographics in Fig. \ref{fig1}. It illustrates the main stages of the construction of the UV completions comparatively, with schematic drawings for the $\Phi$-potential and $\Gamma$-action. 

On the left, the Poincare-conserving vector boson mass term is first promoted to a quartic interaction of the vector boson $V_\mu$ and a scalar $\Phi$ (in spurion status) as in the equation (\ref{MV-map}). Next, this quartic interaction is completed into a gauge-invariant $\Phi$ kinetic term (in Higgs field status) as in (\ref{MV-map2}). These two stages are described by arrows in Fig. \ref{fig1}  with reference to their equation numbers in  Sec. \ref{Gauge Symmetry Restoration at the Extremum of Energy}. The scalar $\Phi$ comes into being through the promotion of $M_\wp$ to $\Phi$ as in the map (\ref{MV-map}). This  equivalence between them, denoted as $M_\wp\rightarrow \Phi$ at the point $\blacksquare$ on the potential energy curve, evolves to the point $\blacktriangle$ at which  the  $\Phi$-potential in (\ref{poten}) attains its minimum at $\Phi=0$ provided that $\mu_\Phi^2 >0$. It is in this sense that $\Phi$ dynamics restores  gauge symmetry and gives meanwhile a UV completion of the effective QFT. 

On the right, the Poincare-breaking gauge boson mass term is first promoted to an interaction term between the gauge boson $V_\mu$ and flat spacetime affine curvature ${\mathbb{R}}({\overline{\Gamma}})$ (in spurion status) as in the equation (\ref{MV-map-new}). Next, this flat spacetime interaction term is carried into curved spacetime with affine curvature ${\mathbb{R}}(\Gamma)$ and metrical curvature $R({}^g\Gamma)$ (in field status)  as in the equation (\ref{MV-map-new-curv}). These two stages are shown with arrows in Fig. \ref{fig1} with reference to their equation numbers in  Sec. \ref{Effective QFT}. Following the discussion  in  Sec. \ref{Effective QFT in Curved Spacetime} of why curvature terms cannot simply be added to the effective QFT, it is shown that the gauge sector is the place where the metrical curvature can arise via general covariance (\ref{covariance}), and a trivial regularization of the flat spacetime affine spurion action (\ref{Sp-curv0}) in Sec. \ref{Gauge Symmetry Restoration at the Extremum of Action} leads to the 
the particular form in (\ref{MV-map-new-curv}). The metric-affine action in (\ref{eff-ac-curv}) is extremized at the affine connection in (\ref{expand-conn}) and this solution leads to restoration of gauge symmetries as in (\ref{reduce-gauge}) and emergence of general relativity as in (\ref{reduce-nongauge}). This action extremization is equivalent to evolution of the affine curvature from its UV value at the point $\bigstar$ (as in equations (\ref{MV-map-new}) or (\ref{MV-map-new-curv})) to the point $\blacklozenge$
at which it attains its IR value of ${\mathbb{R}}(\Gamma)\approx R({}^g\Gamma)$. It is in this sense that at the extremum of the $\Gamma$-action gauge symmetries get restored and general relativity emerges as a quantum effect. 

In summary, as evinced by the infographics in Fig. \ref{fig1},  UV completions of effective QFTs with Poincare-conserving and Poincare-breaking UV cutoffs proceed in parallel with similar phases. The UV completion is realized by Higgs scalars in the former and affine curvature in the latter. It turns out that the gauge and Poincare properties of the UV cutoff have an important impact on the beyond-the-QFT physics needed for UV completion. Characteristic signatures are set of scalar bosons for the Poincare-conserving UV cutoff, and quadratic curvature gravity plus naturally-coupled new particles for the Poincare-breaking UV cutoff.

\section{Conclusion}
\label{Conclusion}
In the present work, a systematic study has been performed of the UV cutoff by grounding on its Poincare properties. The goal is to complete the QFT in the UV with guidance from the Poincare structure of the UV cutoff. The Poincare-conserving and Poincare-breaking UV cutoffs have been analyzed separately with separate examinations of the gauge-conserving and gauge-breaking cases.  Sec. \ref{Poincare-Conserving UV Cutoff} was devoted to the Poincare-conserving UV cutoff. In Sec. \ref{Gauge-Conserving UV Cutoff}, gauge-conserving UV cutoff, main substance of the naturalness criterion, was found to destabilize the light scalars like the Higgs field. In Sec. \ref{Gauge Symmetry Restoration at the Extremum of Energy}, on the other hand, gauge symmetries broken explicitly by the UV cutoff (vector boson mass) were found to get restored by promoting the UV cutoff to appropriate Higgs scalars. 

Sec. \ref{Poincare-Breaking UV Cutoff} was devoted to the Poincare-breaking UV cutoff. In Sec. \ref{Detached Regularization},  detached regularization was introduced as a new regularization framework in which power-law and logarithmic divergences could be analyzed and structured independently. In Sec. \ref{Effective QFT}, flat spacetime effective QFT was formed in the detached regularization with classifications of the power-law and logarithmic divergences. In Sec. \ref{Gauge Symmetry Restoration at the Extremum of Action}, it was shown that the UV cutoff could be consistently promoted to flat spacetime affine curvature in the same philosophy as the promotion of the UV cutoff to Higgs field in Sec. \ref{Gauge-Breaking UV Cutoff}. In Sec. \ref{Effective QFT in Curved Spacetime}, flat spacetime effective QFT was carried into spacetime of a curved metric, with curvature terms arising only in the gauge sector. In Sec. \ref{Affine Dynamics: From UV Cutoff to IR Curvature}, affine curvature was integrated  out the via the affine dynamics and it was found that at the extremum of the affine action gauge symmetries got restored as in (\ref{reduce-gauge}) and gravity (quadratic curvature gravity) emerged as in (\ref{reduce-nongauge}). The salient physics implications listed in Sec. \ref{Salient Properties of the Emergent Gravity} revealed important properties ranging from  existence of new particles beyond the known ones to stability of the light scalars thanks to allowance to naturally weak couplings with the said new particles. In Sec. \ref{Poincare-Conserving vs Poincare-Breaking UV Cutoffs},  a comparative discussion was given of the two completion mechanisms with guidance from the infographics in Fig. \ref{fig1}.

The main lesson from this study is that the Poincare and gauge structures of the UV cutoff have a big say about the likely UV completions of the effective QFT. It turns out that gauge symmetry restoration (in the philosophy of the  Higgs mechanism) plays a detrimental role in specifying the completions. Symmergent gravity, as revealed by its rich gravitational and field-theoretic structure in  Sec. \ref{Poincare-Breaking UV Cutoff}, has the potential to affect various observables and processes in collider, astrophysical and cosmological settings. These opportune observables, partly discussed in Sec. \ref{Salient Properties of the Emergent Gravity}, serve as testbeds for probing and disentangling the symmergent effects.   

\section*{Acknowledgements}
The author thanks to Canan Karahan and Ozan Sarg{\i}n for discussions on bulk and boundary terms in the gauge sector, Ali {\"O}vg{\"u}n and Beyhan Puli{\c c}e on the effects of quadratic curvature term in black hole environments, and {\.I}. {\c C}imdiker, Reggie Pantig, Beyhan Puli{\c c}e and Ali {\"O}vg{\"u}n on the gravitational structure of the potential energy. He thanks also to participants of DICE 2022 Conference for fruitful conversations and discussions, and  acknowledges the contribution of the COST Action
CA21106 - COSMIC WISPers in the Dark Universe: Theory, astrophysics and experiments (CosmicWISPers). The author is grateful to conscientious reviewer for their constructive criticisms, comments and suggestions, which greatly improved the paper in various aspects (including a draft for the second, third and fourth paragraphs in the Introduction). {\bf Note:} The ${\mathcal{O}}(1/M_{Pl}^2)$ subleading terms in (\ref{expand-conn}) and (\ref{expand-curv})  are corrected. This correction does not change any of the results and conclusions. The author thanks Canan Karahan for pointing it out.

\end{document}